\documentclass[aps,twocolumn,showpacs,superscriptaddress]{revtex4-1} 
\usepackage[english]{babel}
\usepackage{graphicx,dcolumn,bm,amssymb,amsmath,amsfonts,amsthm,latexsym}
\usepackage{float,subfig,slashed,hyperref}
\hypersetup{
  linktocpage  = true,
  colorlinks   = true,
  urlcolor     = black,
  linkcolor    = black,
  citecolor    = blue
}

\def\sintcnq{\int \!\!\!\!\!\!\!\!\!\sum_{\ \ cnq}\ }

\def\sintmp{\int \!\!\!\!\!\!\!\!\!\sum_{\ \ mp}\ }

\begin{document}

\title{Relation between the continuum threshold and the Polyakov loop with the QCD deconfinement transition}

\author{J.P.~Carlomagno}
\affiliation{IFLP, CONICET $-$ Dpto.\ de F\'{\i}sica, Universidad Nacional de La Plata, C.C. 67, 1900 La Plata, Argentina}
\affiliation{CONICET, Rivadavia 1917, 1033 Buenos Aires, Argentina}
\author{M.~Loewe}
\affiliation{Instituto de F\'isica, Pontificia Universidad Cat\'olica de Chile, Casilla 306, San\-tia\-go 22, Chile}
\affiliation{Centre for Theoretical and Mathematical Physics and Department of Physics,
University of Cape Town, Rondebosch 7700, South Africa}
\affiliation{Centro Cient\'{\i}fico-Tecnol\'ogico de Valpara\'{\i}so, Casilla 110-V, Valpara\'{\i}so, Chile}     


\begin{abstract}
Using vector and axial-vector correlators within finite energy sum rules with inputs from a chiral quark model, coupled to the Polyakov loop, with nonlocal vector interactions, we extend our previous work to confirm the equivalence between the continuum threshold $s_0$ and the trace of the Polyakov loop $\Phi$ as order parameters for the deconfinement transition at finite temperature $T$ and quark chemical potential $\mu$.
The obtained results are in agreement with our initial conclusion, where we showed that $s_0(T,\mu)$ and $\Phi(T,\mu)$ provide the same information for the QCD deconfinement transition.
\end{abstract}


\pacs{
	25.75.Nq, 
	12.39.Fe, 
	11.55.Hx  
	}
\maketitle


\section{Introduction}
\label{intro}
 
Since quark color charge in a medium is screened due to density and temperature effects, Quantum Chromodynamics (QCD) predicts that at very high temperatures ($T \gg \Lambda_{\rm QCD}$) and low baryon densities,  matter appears in the form of a plasma of quarks and gluons~\cite{Fukushima:2010bq}.  

If one of those variables increases beyond a certain critical value, the interactions between quarks no longer confine them inside hadrons.
This is usually referred to as the deconfinement phase transition.
Simultaneously~\cite{Bazavov:2016uvm}, at small densities another transition takes place, the chiral restoration.
For high values of chemical potential, these two transitions can arise at different critical temperatures. 
The result will be a quarkyonic phase, where the chiral symmetry is restored but the quarks and gluons remain confined.

In the confined region, QCD is strongly coupled and coupling-constant expansions become inapplicable. 
At finite density, lattice QCD (lQCD) methods based on large-scale Monte Carlo simulations are also not applicable because lQCD has the {\em sign problem}~\cite{Splittorff:2007ck,Aarts:2015tyj}. 

Therefore, predictions of the transition features at regions that are not accessible through lattice techniques arise from effective theories, like for instance, the nonlocal Polyakov$-$Nambu$-$Jona-Lasinio (nlPNJL) models (see Ref.~\cite{Carlomagno:2018tyk} and references therein), where quarks interact through covariant nonlocal chirally symmetric couplings in a background color field.
These approaches, which can be considered as an improvement over the local model~\cite{Meisinger:1995ih,Fukushima:2003fw,Megias:2004hj,Ratti:2005jh,Roessner:2006xn,Mukherjee:2006hq,Sasaki:2006ww}, offer a common framework to study both chiral restoration and deconfinement transitions for hadronic systems at finite temperature and/or chemical potential (see e.g.\ Refs.~\cite{GomezDumm:2001fz,GomezDumm:2004sr,Hell:2008cc,Radzhabov:2010dd,Contrera:2010kz,Hell:2011ic,Carlomagno:2013ona,Carlomagno:2018tyk}).
In fact, the nonlocal character of the interactions arises naturally in the context of several successful approaches to low-energy quark dynamics~\cite{Schafer:1996wv,Roberts:1994dr,Roberts:2000aa}, and leads to a momentum dependence in the quark propagator that can be made consistent with lQCD results.
 
In this work we use a nlPNJL model with vector and axial-vector interactions~\cite{Carlomagno:2019yvi}. 
In addition to the standard scalar and pseudoscalar quark-antiquark currents, we consider couplings between vector and axial-vector nonlocal currents, satisfying proper QCD symmetry requirements.

In order to study the properties of chiral and deconfinement phase transitions it has been customary to study the behavior of corresponding order parameters as functions of the temperature and quark chemical potential, namely the quark anti-quark chiral condensate $\langle\bar{q} q\rangle$ and the trace of the Polyakov loop (PL) $\Phi$, respectively. 

In addition to $\Phi$, another phenomenological QCD deconfinement parameter that has been introduced in the literature~\cite{Bochkarev:1986es} is the continuum threshold $s_0$, for the onset of perturbative QCD (PQCD) in hadronic spectral functions.
Around this energy, and at zero temperature, the resonance peaks in the spectrum are either no longer present or become very broad. 

The natural framework to determine $s_0$ has been that of QCD sum rules~\cite{Ayala:2016vnt}. 
This quantum field theory framework is based on the operator product expansion (OPE) of current correlators at short distances, extended beyond perturbation theory, and on Cauchy's theorem in the complex $s$-plane. 

In this article we reconsider the light-quark axial-vector channel with an improved hadronic spectral function involving the $a_1(1260)$ resonance~\cite{Dominguez:2012bs}, in addition to the already considered pion pole approximation~\cite{Carlomagno:2016bpu}. We also include the vector channel with a $\rho$-meson saturated spectral function.

Within this theoretical framework, using finite energy sum rules with inputs  for the spectral functions obtained from a nonlocal quark model (masses, decay constants and widths), we compare the thermal behavior of $s_0$ and $\Phi$ in both channels, at zero and finite chemical potential.

The paper is organized as follows.
In the next section, we briefly review the finite energy sum rules (FESR) program for the light-quark vector and axial-vector channel.
In Sect.~\ref{thermo} we present the general formalism for a finite temperature and density system within the nlPNJL effective model. 
The numerical and phenomenological analyses at zero and finite density are included in Sect.~\ref{results}. 
Finally, in Sect.~\ref{finale} we summarize our results and present the conclusions.


\section{Finite energy sum rules}
\label{fesr}

Within the formalism of finite energy sum rules~\cite{Ayala:2016vnt} we extend our previous work~\cite{Carlomagno:2016bpu}, reconsidering the light-quark axial-vector channel with an improved hadronic spectral function involving the $a_1(1260)$ resonance, in addition to the already considered pion pole approximation. 
Moreover, we also include in this analysis the vector channel with a $\rho$-meson saturated spectral function

We begin considering the current-current correlation function
\begin{equation}
\Pi_{\mu\nu} (q^{2})   = i \, \int\; d^{4} \, x \, e^{i q x} \,
\langle 0|T( J_{\mu}(x)   J_{\nu}^{\dagger}(0))|0\rangle \,,\label{correlator}
\end{equation}
where $J(x)$ is a local quark current.
Let us consider, for our study, the correlator of light-quark vector and axial-vector currents
\begin{eqnarray}
&\Pi_{\mu\nu}^V (q^{2})& = i \int d^{4} \, x \, e^{i q x} \,
\langle 0|T(V_{\mu}(x) V_{\nu}^{\dagger}(0))|0 \rangle \nonumber \\
&=& (-g_{\mu\nu}\, q^2 + q_\mu q_\nu) \, \Pi_0^V(q^2)  \; ,\nonumber \\ 
&\Pi_{\mu\nu}^A (q^{2})& = i \int d^{4} \, x \; e^{i q x} \; 
\langle 0|T( A_{\mu}(x) A_{\nu}^{\dagger}(0))|0 \rangle \nonumber \\
&=& (-g_{\mu\nu}\, q^2 + q_\mu q_\nu) \, \Pi_0^A(q^2) + q_\mu q_\nu\, \Pi_1^A(q^2)  \; ,
\end{eqnarray}
where $A_\mu(x) = : \bar{d}(x) \gamma_\mu \, \gamma_5 u(x):$ is the charged axial-vector current, $V_\mu(x) = \frac{1}{2}[: \bar{u}(x) \gamma_\mu \, u(x) - \bar{d}(x) \gamma_\mu \, d(x):]$ is the electric charge neutral conserved vector current in the chiral limit, and $q_\mu = (\omega, \vec{q})$ is the four-momentum carried by the current. 
The transverse parts $\Pi_0^{A,V}$ are related to the vector meson resonances, whereas the longitudinal axial contribution $\Pi_1^A$ correspond to the pion pole.
Since we are working with non-strange current correlators, there is no longitudinal term in the vector channel~\cite{Dominguez:2018zzi}.

In addition, through Operator Product Expansion of current correlators at short distances~\cite{Shifman:1978bx,QCDSRreview}, one has
\begin{equation}
\Pi^{A,V}(q^2)|_{\mbox{\scriptsize{QCD}}} = C_0 \, \hat{I} + \sum_{N=1} \frac{C_{2N} (q^2,\mu^2)}{(-q^2)^{N}} \langle \hat{\mathcal{O}}_{2N} (\mu^2) \rangle {\Bigr\vert}_{A,V} \;, \label{OPE}
\end{equation}
where $\langle \hat{\mathcal{O}}_{2N} (\mu^2) \rangle \equiv \langle0| \hat{\mathcal{O}}_{2N} (\mu^2)|0 \rangle$, $\mu^2$ is a renormalization scale, the Wilson coefficients $C_N$ depend on the Lorentz indexes and quantum numbers of the currents, and on the local gauge invariant operators ${\hat{\mathcal{O}}}_N$ built from the quark and gluon fields in the QCD Lagrangian. 
These operators are ordered by increasing dimensionality and the Wilson coefficients are calculable in PQCD. 
The unit operator $\hat{I}$ has dimension $d\equiv 2 N =0$, and $C_0$ stands for the purely perturbative contribution. 

On the other hand, using Cauchy's theorem in the complex squared energy $s$-plane, we obtain the quark-hadron duality
\begin{align}
\frac{1}{\pi}\int_{0}^{s_0} ds\ s^N\ \mbox{Im}\  \Pi^{A,V} (s)|_{\mbox{\tiny{HAD}}} = \nonumber \\
   -\frac{1}{2\pi i}\oint_{C(|s_0|)}ds\ s^N\ & \Pi^{A,V} (s)|_{\mbox{\tiny{QCD}}} \;,
\label{disprel}
\end{align}
where the radius of the circle $s_0$ is large enough for QCD and the OPE to be used on the circle. 
Using the OPE, Eq.~(\ref{OPE}), the finite energy sum rules at finite temperature $T$ and chemical potential $\mu$ become~\cite{Bochkarev:1986es,Dominguez:1994re,Ayala:2011vs}
\begin{align}
& (-)^{N-1} C_{2N} \langle {\mathcal{\hat{O}}}_{2N}\rangle {\Bigr\vert}_{A,V} 
= 4 \pi \int_0^{s_0(T,\mu)} ds\, s^{N-1}  \nonumber \\
&\times \Big[ {\mbox{Im}}\ \Pi^{A,V}(s,T,\mu)|_{\mbox{\tiny{HAD}}}
- \, {\mbox{Im}}\ \Pi^{A,V}(s,T,\mu)|_{\mbox{\tiny{PQCD}}} \Big] \;.
\label{FESR}
\end{align}

For $N=1$, the dimension $d=2$ term in the OPE does not involve any condensate, as it is not possible to construct a gauge invariant operator of such dimension from the quark and gluon fields. 
Moreover, there is no evidence for such a term for low values of $T$~\cite{Dominguez:1999xa, Dominguez:2006ct, Megias:2009ar}. 

The dimension $d=4$ term, a renormalization group invariant quantity, is given in the chiral limit, for the vector and axial sector by 
\begin{eqnarray}
\frac{1}{2}\ C_4 \langle \hat{\mathcal{O}}_{4}  \rangle \ \big{\vert}_V &=&
C_4 \langle \hat{\mathcal{O}}_{4}  \rangle \ \big{\vert}_A  =
\frac{\pi^2}{6} \langle \frac{\alpha_s}{\pi} G_{\mu \nu}G^{\mu \nu} \rangle.
\label{C4}
\end{eqnarray}

As was mentioned previously, there is a difference in the currents involved in vector and axial processes, i.e. the former involves electrically neutral currents, while the latter involves electrically charged currents. 
Then, the normalization of these vector current correlators differs by a factor two.

The leading power correction of dimension $d=6$ is the four-quark condensate, which in the vacuum saturation approximation~\cite{QCDSRreview} becomes
\begin{eqnarray}
C_6 \langle \hat{\mathcal{O}}_{6}  \rangle \ \big{\vert}_V \propto
C_6 \langle \hat{\mathcal{O}}_{6}  \rangle \ \big{\vert}_A  \propto
\alpha_s \,|\langle \bar{q} q \rangle|^2\;,
\label{C6}
\end{eqnarray}
which is channel dependent and has a very mild dependence on the renormalization scale.
This approximation has no solid theoretical justification, other than its simplicity.

In the static limit ($\vec{q} \rightarrow 0$), and for finite $T$ and $\mu$, the spectral function in PQCD, $\Pi^{A,V}(\omega^2, T, \mu)|_{\mbox{\tiny{PQCD}}}$ (to simplify the notation we shall omit the $s$, $T$ and $\mu$ dependence), is given by
\begin{align}
 & {\mbox{Im}}\Pi^A|_{\mbox{\tiny{PQCD}}} =
   {\mbox{Im}}\Pi^V|_{\mbox{\tiny{PQCD}}} = \nonumber \\ 
 & \frac{1}{4\pi}\left[1-n_+\left(\frac{\sqrt{s}}{2}\right) 
   -n_-\left(\frac{\sqrt{s}}{2}\right)\right] \nonumber \\
 &  -\frac{2}{\pi} \;T^2 \;\delta (s)\; \left[
   {\mbox{Li}}_2(-e^{\mu/T})
   + {\mbox{Li}}_2(-e^{-\mu/T})\right],
\label{pertQCD}
\end{align}
where ${\mbox{Li}}_2(x)$ is the dilogarithm function, $s=\omega^2$, and
\begin{equation}
   n_\pm(x)=\frac{1}{e^{(x\mp \mu)/T}+1}
\label{F-D}
\end{equation}
are the Fermi-Dirac thermal distributions for particles and antiparticles, respectively. 

\begin{figure}[h]
\begin{center}
\centering
\includegraphics[width=0.49\textwidth]{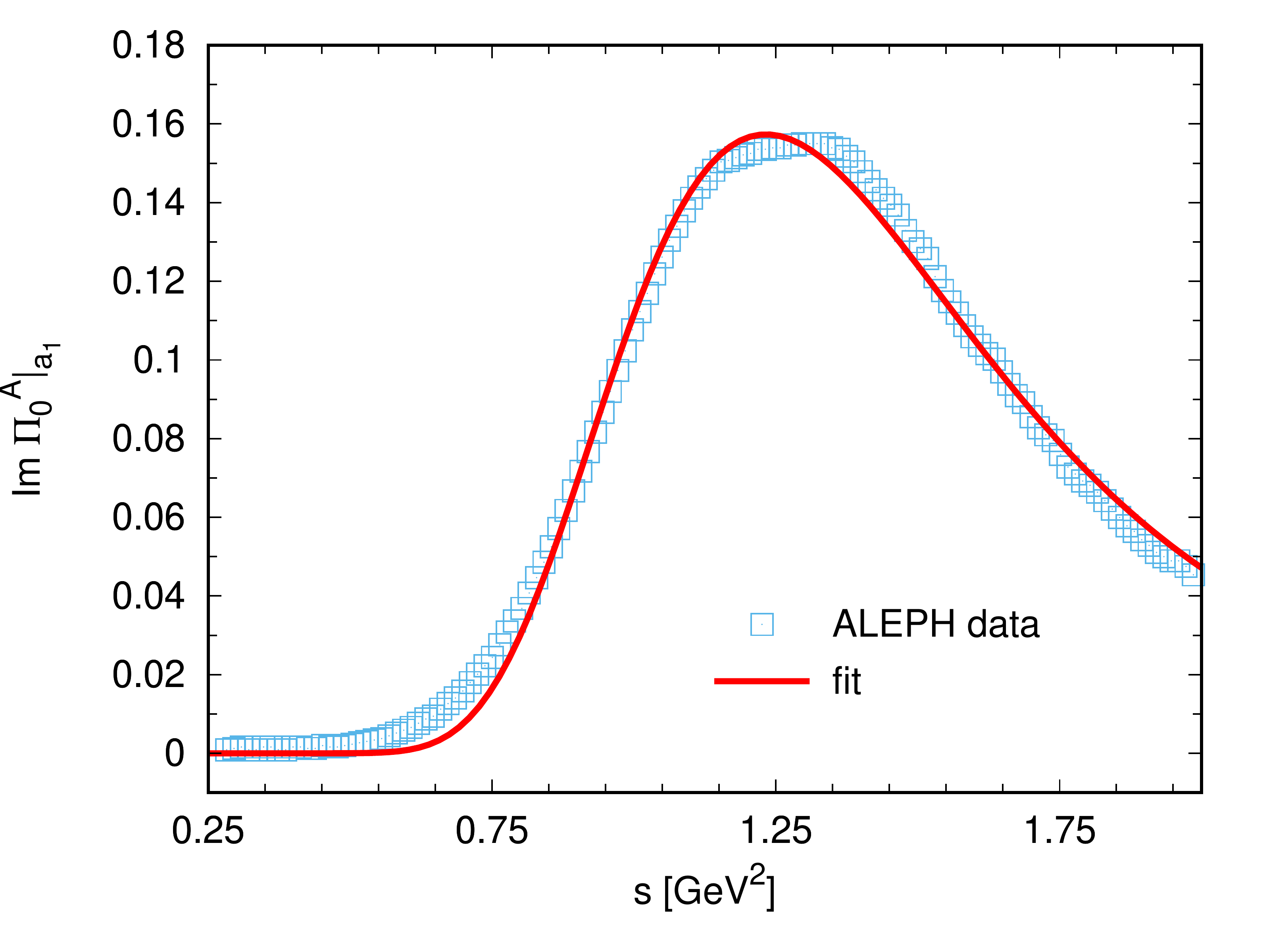}
\end{center}
\caption{
Fit to ALEPH data in the axial-vector channel~\cite{Schael:2005am} up to $s \simeq 2\, {\mbox{GeV}}^2$. Errors are represented by the size of the data points.}
\label{fig:a1}
\end{figure}

In the chiral limit, the axial-vector hadronic sector in the spectral function can be approximated by the pion pole, followed by the $a_1(1260)$ resonance
\begin{equation}
{\mbox{Im}}\, \Pi^A|_{\mbox{\scriptsize{HAD}}} = 2 \,\pi\, f_\pi^2 \;\delta(s) + {\mbox{Im}}\, \Pi^A|_{a_1}\;, \label{HADA}
\end{equation} 
where $f_\pi$  is the pion decay constant, and a fit in the resonance region ($s < 2\, {\mbox{GeV}}^2$) to the ALEPH data quoted in Ref.~\cite{Schael:2005am}, gives
\begin{equation}
{\mbox{Im}} \;\Pi^A|_{a_1} = \pi\ \alpha \, f_{a_1} \; \exp{\, \left[- \left(\frac{s - \delta\ m_{a_1}^2}{\Gamma_{a_1}^2 + \beta\ s}\right)^2\right]}\;, \label{A1}
\end{equation}
with $\alpha = 0.1136 \pm 0.0002$, $\beta = 0.302 \pm 0.001$ and $\delta = 0.8501 \pm 0.0006$.
The mass, width and decay constants of the $a_1$ meson, $m_{a_1}$, $\Gamma_{a_1}$ and $f_{a_1}$, respectively, are obtained from a two flavor nonlocal PNJL model with vector interactions~\cite{Carlomagno:2019yvi}, that will be described in Sect.~\ref{thermo}.
This fit together with the ALEPH data are shown in Fig.~\ref{fig:a1}.

In our previous work~\cite{Carlomagno:2016bpu} we only considered for the axial spectral function the pion pole. 
Therefore, incorporating the $a_1$ resonance improves the approximation and constitutes the best possible approach within the present theoretical advances.
 
For the vector sector, as usual~\cite{Ayala:2012ch,Ayala:2013vra,Ayala:2014rka}, we will assume $\rho$-meson saturation of the spectral function in terms of a Breit-Wigner resonance.
This parametrization has been normalized such that its area is equal to the area under a zero-width expression~\cite{Ayala:2012ch}.
In addition, we will consider a contribution due to the coupling of the vector current to two pions in the thermal bath, the so called scattering term, and is given by
\begin{equation}
{\mbox{Im}}\, \Pi^V|_{S} = \frac{2}{3\pi}\delta(s)\int_0^\infty y\, n_B\left(\frac{y}{T}\right)dy = \frac{\pi}{9}\delta(s) T^2 \;,
\end{equation}
where $n_B(z) = 1/(e^{z}-1)$ is the Bose thermal function.	

Therefore, we have
\begin{equation}
\frac{1}{2}\ {\mbox{Im}} \Pi^V|_{HAD} = f_\rho^2 \frac{m_\rho^3 \Gamma_\rho}{\left(s - m_\rho^2\right)^2 + m_\rho^2 \Gamma_\rho^2} + \frac{1}{8\pi}\ {\mbox{Im}}\, \Pi^V|_{S}\ , \label{HADV}
\end{equation}
where as before, $m_\rho$, $\Gamma_\rho$ and $f_\rho$ are the $\rho$ mass, width and decay constants, respectively, also calculated within the nlPNJL model.                     


\section{Thermodynamics at finite density}
\label{thermo}

In this section we present the formalism of a two-flavor quark model coupled to the Polyakov loop, that includes nonlocal vector and axial-vector quark-antiquark currents, in addition to the standard scalar and pseudo-scalar nonlocal interactions~\cite{Carlomagno:2019yvi}.
The corresponding Euclidean effective action is given by~\cite{Villafane:2016ukb}
\begin{align}
 S_{E} &= \int d^{4}x\ \bigg\{ \bar{\psi}(x)\left( -i\gamma_{\mu}D_{\mu} +\hat{m}\right)  \psi(x) \nonumber \\
& - \frac{G_{S}}{2} \Big[ j_S(x)j_S(x) + j_P^a(x)j_P^a(x) + j_M(x)j_M(x)\Big] \nonumber \\
& - \left. \frac{G_{V}}{2} \Big[ j_V^{\mu a}(x)j_{V \mu}^a(x) + j_A^{\mu a}(x)j_{A \mu}^a(x)\Big] \right. \nonumber \\
& \left. -\ \frac{G_{0}}{2} j_V^{\mu 0}(x)j_{V \mu}^0(x) - \frac{G_{5}}{2} j_A^{\mu 0}(x)j_{A \mu}^0(x)  \right. \nonumber \\
& \left. +\ {\cal U}\,(\Phi[A(x)]) \right\rbrace \ , 
\label{action}%
\end{align}
where $\psi$ is the $N_{f}=2$ fermion doublet $\psi\equiv(u,d)^T$, and $\hat{m}={\rm diag}(m_{u},m_{d})$ is the current quark mass matrix. In what follows we consider isospin symmetry, $m_{u}=m_{d}=m$. 
The fermion kinetic term in Eq.~(\ref{action}) includes a covariant derivative $D_\mu\equiv\partial_\mu - iA_\mu$, where $A_\mu = g\, G^\mu_a \lambda^a/2$, with $G^\mu_a$ the SU(3) color gauge fields.
The nonlocal currents are given by
\begin{align}
j_S(x)  &  =\int d^{4}z\ {\cal G}(z)\ \bar{\psi}\left(  x+\frac{z}{2}\right)
\ \psi\left(  x-\frac{z}{2}\right)  \ ,\nonumber\\
j_P^a(x)  &  =\int d^{4}z\ {\cal G}(z)\ \bar{\psi}\left(  x+\frac{z}{2}\right)
\ i\gamma_{5}\tau^a\ \psi\left(  x-\frac{z}{2}\right)  \ ,\nonumber\\
j_M(x)  &  =\int d^{4}z\ {\cal F}(z)\ \bar{\psi}\left(  x+\frac{z}{2}\right)
\ \frac{i {\overleftrightarrow{\rlap/\partial}}}{2\ \kappa_{p}}
\ \psi\left(  x-\frac{z}{2}\right)\ ,\nonumber\\
j_V^{\mu a}(x)  &  =\int d^{4}z\ {\cal G}(z)\ \bar{\psi}\left(  x+\frac{z}{2}\right)
\ \gamma^\mu \tau^a \ \psi\left(  x-\frac{z}{2}\right)  \ ,\nonumber\\
j_A^{\mu a}(x)  &  =\int d^{4}z\ {\cal G}(z)\ \bar{\psi}\left(  x+\frac{z}{2}\right)
\ \gamma^\mu \gamma_5 \tau^a \ \psi\left(  x-\frac{z}{2}\right) ,\nonumber\\
j_V^{\mu 0}(x)  &  =\int d^{4}z\ {\cal G}(z)\ \bar{\psi}\left(  x+\frac{z}{2}\right)
\ \gamma^\mu \psi\left(  x-\frac{z}{2}\right)  \ ,\nonumber\\
j_A^{\mu 0}(x)  &  =\int d^{4}z\ {\cal G}(z)\ \bar{\psi}\left(  x+\frac{z}{2}\right)
\ \gamma^\mu \gamma_5 \ \psi\left(  x-\frac{z}{2}\right) ,
\label{currents}%
\end{align}
where $u(x^{\prime}){\overleftrightarrow{\partial}}v(x)=u(x^{\prime})\partial_{x}v(x)-\partial_{x^{\prime}}u(x^{\prime})v(x)$, $\tau_a$ are the Pauli matrices and the functions ${\cal G}(z)$ and ${\cal F}(z)$ in Eq.~(\ref{currents}) are nonlocal covariant form factors characterizing the corresponding interactions.  
The scalar-isoscalar component of the nonlocal currents will generate a momentum dependent quark mass in the quark propagator, while the ``momentum'' current $j_M(x)$  leads to a momentum-dependent wave function renormalization of the quark propagator, in consistency with lQCD analyses.

To work with mesonic degrees of freedom, the fermionic theory is bosonized in a standard way~\cite{Ripka:1997zb} by considering the corresponding partition function $\mathcal{Z} = \int \mathcal{D}\, \bar{\psi}\mathcal{D}\psi \,\exp[-S_E]$, introducing auxiliary bosonic fields $\sigma_1(x)$, $\sigma_2(x)$ (scalar, related to $j_S(x)$ and $j_M(x)$), $\pi^a(x)$ (pseudoscalar), $v^a_\mu(x)$ (vector) and $a^a_\mu (x)$ (axial vector) and integrating out the quark fields.
Details of this procedure can be found in Ref.~\cite{Villafane:2016ukb}.

Since we are interested in deconfinement and chiral restoration transitions, we extend the bosonized effective action to finite temperature $T$ and chemical potential $\mu$. 
This will be done using the standard imaginary time formalism. 
Concerning the gauge fields $A_\mu$, we assume that quarks move in a constant background field $\phi = A_4 = i A_0 = i g\,\delta_{\mu 0}\, G^\mu_a \lambda^a/2$. 
We will work in the so-called Polyakov gauge, in which the matrix $\phi$ is given a diagonal representation $\phi = \phi_3 \lambda_3 + \phi_8 \lambda_8$.

Although at finite chemical potential, the traced PL $\Phi$ and its conjugate $\Phi^\ast$ could in principle have different values, the difference between the $\Phi$ and $\Phi^\ast$ emerges beyond the mean field approximation~\cite{Roessner:2006xn,Ratti:2006wg,Ratti:2007jf}.
Therefore we choose for our model to adopt the usual prescription $\Phi = \Phi^\ast$ (see Ref.~\cite{Carlomagno:2019yvi}) implying that $\phi_8=0$, leaving only $\phi_3$ as an independent variable.
Consequently, the traced Polyakov loop results $\Phi = [ 2 \cos(\phi_3/T) + 1 ]/3$

For the light quark sector the trace of the Polyakov loop turns out to be an approximate order parameter, in the same way the chiral quark condensate is an approximate order parameter for the chiral symmetry restoration.

Thus, in the mean field approximation (MFA), and following the same prescriptions as in Refs.~\cite{GomezDumm:2001fz,GomezDumm:2004sr,Carlomagno:2018tyk}, the thermodynamic potential $\Omega^{\rm MFA}$ at finite temperature $T$ and chemical potential $\mu$ is given by~\cite{Contrera:2012wj}
\begin{equation}
\label{omegareg}
\Omega^{\rm MFA} \ = \ \Omega^{\rm reg} + \Omega^{\rm free} +
\mathcal{U}(\Phi,T) + \Omega_0 \ ,
\end{equation}
where
\begin{widetext}
\begin{align}
\Omega^{\rm reg} &=  \,- \,4\ \sintcnq \ 
\ln \left[ \frac{q_{n,c}^2 + m^2(q_{n,c})}{z^2(q_{n,c})\ (q_{n,c}^2 + m^2)}\right]+
\frac{\bar\sigma_1^2 + \kappa_p^2\; \bar\sigma_2^2}{2\,G_S} - \frac{\bar\omega^2}{2\,G_0} \ , \nonumber \\
\Omega^{\rm free} \ &= \ -4 T \sum_{c=r,g,b} \ \sum_{s=\pm 1} \int \frac{d^3 \vec{q}}{(2\pi)^3}\; \mbox{Re}\;
\ln \left[ 1 + \exp\left(-\;\frac{\epsilon_q + s(\mu + i \phi_c)}{T}
\right)\right]
\ ,
\label{granp}
\end{align}
\end{widetext}
whith the shorthand notation
\begin{equation}
\sintcnq  \equiv \ \sum_{c=r,g,b} T \sum_{n=-\infty}^{\infty}
\int \frac{d^3\vec q}{(2\pi)^3} \ . \nonumber
\end{equation}
The constants $\bar\sigma_{1,2}$ and $\bar{\omega}$ are the mean field values of the scalar fields and the isospin zero vector field.
At nonzero quark densities, the flavor singlet term of the vector interaction develops a nonzero expectation value $\bar{\omega}$, while all other components of the vector and axial vector interactions have vanishing mean fields~\cite{Bratovic:2012qs}.

The mean field values can be calculated by minimizing $\Omega^{\rm MFA}$, while the functions $m(p)$ and $z(p)$ ---momentum-dependent effective mass and wave function renormalization (WFR)--- are related to the nonlocal form factors and the vacuum expectation values of the scalar fields by
\begin{eqnarray}
m(p) &=& z(p)\, \left[ m\, +\, \bar \sigma_1\, g(p)\right] \ ,\nonumber\\
z(p) &=& \left[ 1\,-\,\bar \sigma_2 \,f(p)\right]^{-1}\ ,
\label{mz}
\end{eqnarray}
where $g(p)$ and $f(p)$ are the Fourier transforms of the form factors in Eq.~(\ref{currents}).
We have also defined
\begin{eqnarray}
q_{n c}^2 = 
\Big[\omega_n  + \phi_c - i \tilde{\mu} \Big]^2 + {\vec{q}}\ \! ^2 \ , 
\end{eqnarray}
the sums over color indices run over $c=r,g,b$, with the color background fields components being $\phi_r=-\phi_g=\phi$, $\phi_b = 0$, $\epsilon_q = \sqrt{\vec{q}^{\;2}+m^2}\;$, and $\omega_n = (2 n +1 )\pi  T$ are the fermionic Matsubara frequencies. 
The vector coupling generates a shifting in the chemical potential as~\cite{Contrera:2012wj}
\begin{equation}
\tilde{\mu} = \mu - g(\bar{q}_{n c})\ z(\bar{q}_{n c})\ \bar{\omega} \ ,
\end{equation}
where 
\begin{equation}
\bar{q}_{n c} = q_{n c}\ \vert_{\bar{\omega}=0} \ .
\end{equation}

The term $\Omega^{\rm reg}$ is the regularized expression with the thermodynamic potential of a free fermion gas $\Omega^{\rm free}$, and finally the last term in Eq.~(\ref{omegareg}) is just a constant fixed by the condition that $\Omega^{\rm MFA}$ vanishes at $T=\mu=0$.

The effective gauge field self-interactions are given by the Polyakov loop potential $\mathcal{U}(\Phi,T)$. 
At finite temperature, it is usual to take for this potential a functional form based on properties of pure gauge QCD. 
Among the most used effective potentials, the Ansatz that provides the best agreement with lQCD results~\cite{Carlomagno:2013ona,Carlomagno:2018tyk} is the polynomial function based on a Ginzburg-Landau Ansatz~\cite{Ratti:2005jh,Scavenius:2002ru}:
\begin{align}
\frac{{\cal{U}}_{\rm poly}(\Phi ,T)}{T ^4} \ = \ -\,\frac{b_2(T)}{2}\, \Phi^2
-\,\frac{b_3}{3}\, \Phi^3 +\,\frac{b_4}{4}\, \Phi^4 \ ,
\label{upoly}
\end{align}
where
\begin{align}
b_2(T) = a_0 +a_1 \left(\dfrac{T_0}{T}\right) + a_2\left(\dfrac{T_0}{T}\right)^2
+ a_3\left(\dfrac{T_0}{T}\right)^3\ .
\label{pol}
\end{align}

Numerical values for the parameters can be found in Table~\ref{tab:setpoly}~\cite{Ratti:2005jh}.
\begin{table}[h]
\begin{center}
\begin{tabular*}{0.35\textwidth}{@{\extracolsep{\fill}} c c c c c c}
\hline
\hline
$a_0$&$a_1$&$a_2$&$a_3$&$b_3$&$b_4$\\
\hline
6.75&-1.95&2.625&-7.44&0.75&7.5\\
\hline
\hline
\end{tabular*}
\caption{\small{Parameter set used for the polynomial Polyakov loop potential, Eq.~(\ref{upoly}).}}
\label{tab:setpoly}
\end{center}
\end{table}

In absence of dynamical quarks, from lattice calculations one expects a deconfinement temperature $T_0 = 270$~MeV.
However, it has been argued that in the presence of light dynamical quarks this temperature scale, a further parameter of the model, should be adequately reduced to about $210$ and $190$~MeV for the case of two and three flavors, respectively, with an uncertainty of about $30$~MeV~\cite{Schaefer:2007pw}. 

In order to fully specify the model under consideration, we proceed to fix the model parameters as well as the nonlocal form factors $g(q)$ and $f(q)$. We consider here Gaussian functions 
\begin{align}
g(q) &= \mbox{exp}\left(-q^{2}/\Lambda_{0}^{2}\right) \ , \nonumber \\
f(q) &= \mbox{exp}\left(-q^{2}/\Lambda_{1}^{2}\right)\ ,
\label{regulators}
\end{align}
which guarantee a fast ultraviolet convergence of the loop integrals and offer a momentum dependence in good agreement with lQCD, and other form factors~\cite{Carlomagno:2013ona,Villafane:2016ukb}.

Once the mean field values are obtained, the behavior of other relevant quantities as functions of the temperature and chemical potential can be determined. 
We concentrate, in particular, on the chiral quark condensate $\langle\bar{q}q\rangle = \partial\Omega^{\rm MFA}_{\rm reg}/\partial m$ and the traced PL $\Phi$, which will be taken as order parameters for the chiral restoration and deconfinement transitions, respectively. 
The associated susceptibilities will be defined as $\chi_{\rm ch}  = \partial\,\langle\bar qq\rangle/\partial T$ and $\chi_{\Phi} = d \Phi / d T$. 

\subsection*{Meson masses}
Meson masses can be obtained from the terms in the Euclidean action that are quadratic in the bosonic fields. 
From the zero temperature action~\cite{Villafane:2016ukb} one can obtain, using the imaginary time formalism, the finite temperature action,
\begin{widetext}
\begin{eqnarray} \label{eq:quad}
&& S_E^{\rm quad} = \dfrac{1}{2}\ \sintmp
		\Big\{ G_{\sigma\sigma}(\nu_m,\vec{p})\, \delta\sigma(p_m)\, \delta\sigma(-p_m)  +
		G_{\sigma^\prime\sigma^\prime}(\nu_m,\vec{p})\, \delta\sigma^\prime(p_m)\,\delta\sigma^\prime(-p_m) +	G_{\pi\pi}(\nu_m,\vec{p}) \, \delta\vec\pi(p_m)\cdot\delta\vec\pi(-p_m)\nonumber\\
&& +\; i\,G_{\pi a}(\nu_m,\vec{p})\Big[p_m^\mu\,\delta\vec{a}_{\mu}(-p_m) \cdot
        \delta\vec{\pi}(p_m)-p_m^\mu\,\delta\vec{a}_{\mu}(p_m)
        \cdot \delta\vec{\pi}(-p_m)\Big]
        +\;G_{vv}^{\mu\nu}(\nu_m,\vec{p})\,\delta\vec{v}_\mu(p_m)\cdot\delta\vec{v}_\nu(-p_m) \nonumber \\
&&		+\;G_{aa}^{\mu\nu}(\nu_m,\vec{p})\,\delta\vec{a}_\mu(p_m)\cdot\delta\vec{a}_\nu(-p_m) 
        +\; G_{vv}^{\mu\nu,0}(\nu_m,\vec{p})\,\delta v^0_{\mu}(p_m)\,\delta v^0_{\nu}(-p_m)  
        + G_{aa}^{\mu\nu,5}(\nu_m,\vec{p})\,\delta a^0_{\mu}(p_m)\,\delta a^0_{\nu}(-p_m) \Big\}\ ,
\label{sequad}
\end{eqnarray}
\end{widetext}
with $p_m\equiv(\nu_m,\vec{p})$.
The functions $G_{ab}(\nu_m,\vec{p})$ are given by finite temperature one-loop integrals arising from the fermionic determinant in the bosonized action, and $\nu_m = 2 m \pi T$ are the bosonic Matsubara frequencies.
Once cross terms have been eliminated, the resulting functions $G_{M}(\nu_m,\vec{p})$ stand for the inverses of the effective thermal meson propagators, in the imaginary time formalism.
The functions $G_{\rho,{\rm a}_1}(\nu_m,\vec{p})$ correspond to the transverse projections of the vector and axial-vector fields. 
Thus the masses of the physical $\rho^0$, $\rho^\pm$ (which are degenerate in the isospin limit) and $a_1$ (the transverse parts of the $\vec{a}_\mu$ fields do not mix with the pions) can be obtained by solving the equation $G_{\rho,{\rm a}_1}(0,-i m_{\rho,{\rm a}_1})\ =\ 0$, where
\begin{eqnarray}
\hspace{-4mm} && G_{\rho \choose {\rm a}_1}(\nu_m,\vec{p}) = \dfrac{1}{G_V}-8\, \sintcnq  
                               h^{2}(q_{nc})\,
                               \dfrac{z(q_{nc}^+)z(q_{nc}^-)}{D(q_{nc}^{+})D(q_{nc}^{-})} \nonumber \\
                         &&\times \left[\dfrac{q_{nc}^{2}}{3}+\dfrac{2(p_m\cdot q_{nc})^{2}}{3p^{2}}-
                               \dfrac{p_m^{2}}{4}\pm m(q_{nc}^{-})m(q_{nc}^{+})\right],
                               \label{grho} 
\end{eqnarray}
with $D(q_{nc}) = q_{nc}^2 + m^2(q_{nc})$ and $q_{nc}^\pm = q_{nc} \pm p_m/2$.
\label{rmass}

In the case of the pseudoscalar sector, from Eq.~(\ref{sequad}) it is seen that there is a mixing between the pion fields and the longitudinal part of the axial vector fields~\cite{Ebert:1985kz,Bernard:1993rz}.
Therefore, after the elimination of the mixing term, the pion mass can be then calculated from $G_{\tilde\pi}(0,-i m_\pi) = 0$, where
\begin{equation}
G_{\tilde{\pi}}(p_m^2)= G_{\pi}(p_m^2)-\dfrac{G_{\pi a}^2(p_m^2)}{L_-(p_m^2)}\,p_m^2\ ,
\end{equation}
with
\begin{eqnarray}
	     && G_{\pi}(p_m^2)\  =  \dfrac{1}{G_S} \, - \, 8\ 
		\sintcnq \ g(q_{nc})^2\,
                	\dfrac{z(q_{nc}^+)z(q_{nc}^-)}{D(q_{nc}^{+})D(q_{nc}^{-})}\nonumber \\
       && \times \left[(q_{nc}^{+}\cdot q_{nc}^-)\,+\,m(q_{nc}^{+})\,m(q_{nc}^{-})\right] \ , \nonumber \\
		&&G_{\pi a}(p_m^2)\ =  \dfrac{8}{p_m^{2}}\, 
		\sintcnq \ g(q_{nc})\,h(q_{nc})\,
                 	\dfrac{z(q_{nc}^+)z(q_{nc}^-)}{D(q_{nc}^{+})D(q_{nc}^{-})}\nonumber \\
       && \times \left[(q_{nc}^{+}\cdot p_m)\,m(q_{nc}^{-})-(q_{nc}^{-}\cdot
                 p_m)\,m(q_{nc}^{+})\right] \ , \nonumber \\
		&&L_{-}(p_m^{2})\  = \dfrac{1}{G_V}-8
		\sintcnq \ h^{2}(q_{nc})\,
                 	\dfrac{z(q_{nc}^+)z(q_{nc}^-)}{D(q_{nc}^{+})D(q_{nc}^{-})}\nonumber \\
       && \times \left[q_{nc}^{2}-\dfrac{2(p_m\cdot q_{nc})^{2}}{p_m^{2}}+\dfrac{p_m^{2}}{4} -
                	 m(q_{nc}^{-})m(q_{nc}^{+})\right] . 
\label{lpm}                 
\end{eqnarray}

\subsection*{Meson decay constants}
The pion weak decay constant $f_\pi$ is given by the matrix elements of axial currents between the vacuum and the
physical one-pion states at the pion pole,
\begin{equation}
\label{eq:fpi} 
\langle 0 \vert J_{A\mu}^a (0) \vert \tilde{\pi}^b(p) \rangle = 
i \, \delta^{ab} \, f_\pi(p^2) \; p_\mu \ .
\end{equation}

On the other hand, the matrix elements of the electromagnetic current $J_{em}$ between the neutral vector meson state and the vacuum determine the vector decay constant $f_\rho$~\cite{Villafane:2016ukb},
\begin{equation}
\label{eq:frho} 
\langle 0 \vert J_{em\ \mu} (0) \vert \rho_\nu (p) \rangle = 
e \, f_\rho(p^2) (g_{\mu\nu} p^2 - p_\mu p_\nu) \ ,
\end{equation}
with $p^2 = -m_\rho^2$, where $e$ is the electron charge.

And finally, the axial-vector decay constant $f_{a_1}$ is defined by the matrix elements of the electroweak charged currents $J_{ew}$ between the axial-vector meson state and the vacuum, at $p^2 = -m_{a_1}^2$, as~\cite{Carlomagno:2019yvi}
\begin{equation}
\label{eq:fa1} 
\langle 0 \vert J_{ew\ \mu} (0) \vert a_{1 \nu}(p) \rangle = 
 f_{a_1}(p^2) (g_{\mu\nu} p^2 - p_\mu p_\nu)\ . 
\end{equation}

In order to obtain these matrix elements within our model, we have to “gauge” the effective action through the introduction of gauge fields, and then we have to calculate the functional derivatives of the bosonized action with respect to the currents and the renormalized meson fields.
In addition, due to the nonlocality of the interaction, the gauging procedure requires the introduction of gauge fields not only through the usual covariant derivative in the Euclidean action, but also through a transport function that comes with the fermion fields in the nonlocal currents (see e.g. Refs.~\cite{Ripka:1997zb,Bowler:1994ir,GomezDumm:2006vz}).

After a lengthy calculation, the decay constants at finite $T$ and $\mu$, are given by
\begin{eqnarray}
f_\pi &=& \dfrac{m_q \, Z_\pi^{1/2}}{m_\pi^2} \left[ F_0 (0,-i m_\pi)
+\dfrac{G_{\pi a}(p_m^2)}{L_-(p_m^2)} \, F_1 (0,-i m_\pi)\right] ,\nonumber\\
f_\rho \ &=& \ \dfrac{Z_\rho^{1/2}}{3\, m_\rho^2}
\, \left[ J_V^{\rm (I)} (0,-i m_\rho) + J_V^{\rm (II)} (0,-i m_\rho)\right] ,\nonumber\\
f_{a_1} \ &=& \ \dfrac{Z_{a_1}^{1/2}}{3\, m_{a_1}^2}
\, \left[ J_A^{\rm (I)} (0,-i m_{a_1}) + J_A^{\rm (II)} (0,-i m_{a_1})\right] ,
\label{efes}
\end{eqnarray}
where the mesons wave function renormalization can be obtained from
\begin{equation}
\label{zpi} Z_M^{-1}= \frac{dG_M(p_m^2)}{dp_m^2}\bigg\vert_{p_m^2=-m_M^2} \ .
\end{equation}

The analytical expressions for $F_0$, $F_1$, $J_{V,A}^{\rm (I, II)}$ and the thermal behavior of $f_\pi$, $f_v$ and $f_a$ can be found in Ref.~\cite{Carlomagno:2019yvi}.

In absence of vector meson fields, the mixing term $F_1$ in  $f_\pi$ vanishes, and the expression reduces to that quoted in Ref.~\cite{Noguera:2008cm}.

\begin{figure}[H]
\begin{center}
\includegraphics[width=0.35\textwidth]{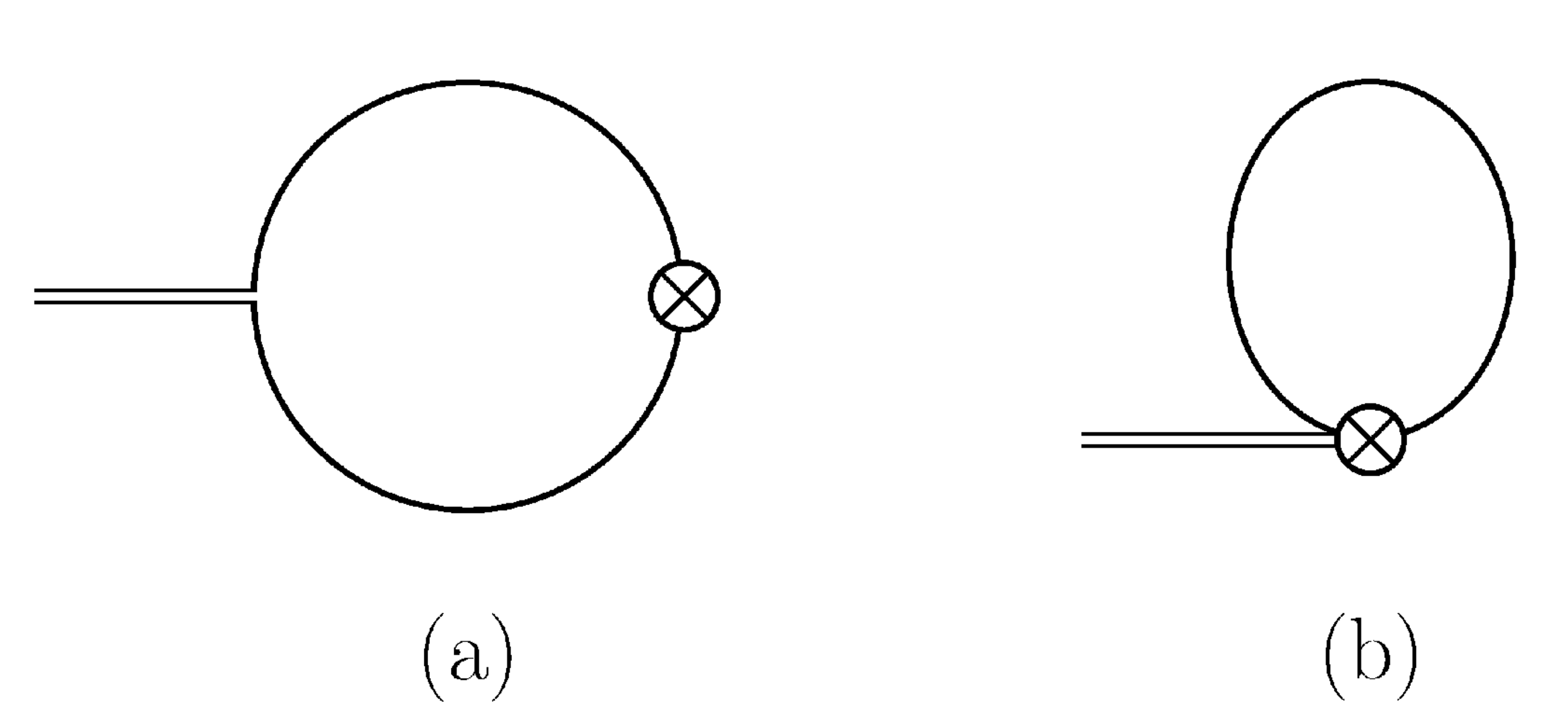}
\end{center}
\caption{Diagrammatic representation of the contributions to the weak decay constants. The cross represents the axial current vertex. (a) Two-vertex diagram. (b) Tadpolelike contribution.}
\label{fig:diagefes}
\end{figure}

The resulting one-loop contributions are diagrammatically schematized in Fig.~\ref{fig:diagefes}. 
Tadpole-like diagrams, which are not present in the local PNJL model, arise from the occurrence of gauge fields.

\subsection*{Decay widths}
In general, various transition amplitudes can be calculated by expanding the bosonized action to higher orders in meson fluctuations. 
The decay amplitudes $\mathcal{A}_\rho(\rho \to \pi\pi)$ and 
$\mathcal{A}_{a_1}(a_1 \to \rho\pi)$ are obtained by calculating the corresponding  functional derivatives of the effective action.

For the vector sector, only the transverse piece contributes to $\rho\to\pi\pi$ decay~\cite{Villafane:2016ukb}, while for the axial-vector sector  both transverse and longitudinal parts contribute to the $a_1\to\rho\pi$ decay~\cite{Carlomagno:2019yvi}.

In order to study the thermal dependence of these decay widths, it is necessary to modify the two-body phase space to include finite temperature effects.

Following Refs.~\cite{Weldon:1991ei,Tanabashi:2018oca}, the decay of a particle at rest of mass $M$, into particles of masses $m_1$ and $m_2$ in equilibrium with the heat bath, is given by
\begin{eqnarray}
\label{gammaofT}
&&\Gamma_{M\to m_1 m_2}\ \big\vert_{p=0} = \frac{\vert \mathcal{A}_M \vert^2}{32 \pi M}\ \nonumber \\
&& \times \sqrt{\left(1-\frac{(m_1 + m_2)^2}{M^2}\right)
\left(1-\frac{(m_1 - m_2)^2}{M^2}\right)}\  \nonumber \\
&& \times \frac{\exp{\left[\frac{1}{2T}M\right]}}{\cosh{\left[\frac{1}{2T}M\right]}-\cosh{\left[\frac{1}{2T}\frac{(m_1-m_2)(m_1+m_2)}{M}\right]}} \ ,
\end{eqnarray}
where $\mathcal{A}_M$ is evaluated within the effective model.

For the decays $\rho \rightarrow \pi \pi$ ($M=m_\rho$, $m_1=m_2=m_\pi$) and $a_1 \rightarrow \rho \pi$ ($M=m_{a_1}$, $m_1=m_\rho$, $m_2=m_\pi$) we have
\begin{eqnarray}
\label{eq:amplitude}
\vert \mathcal{A}_\rho \vert^2 &=& 
\dfrac{m_\rho^2}{3} \bigg(1-\dfrac{4m_\pi^2}{m_\rho^2}\bigg) g_{\rho\pi\pi}^2  \ , \nonumber\\
\vert \mathcal{A}_{a_1} \vert^2 &=& 
2\,g_{a\rho\pi}^2 + 
\dfrac{1}{16\,m_\rho^2m_{a_1}^2} \Big\{2\,g_{a\rho\pi}(m_{a_1}^2-m_\pi^2+m_\rho^2) \nonumber\\
	& & \hspace{-0.5cm} + f_{a\rho\pi}\big[m_{a_1}^4-2\,m_{a_1}^2(m_\rho^2+m_\pi^2)+(m_\rho^2-m_\pi^2)^2\big]\Big\}^2 .\nonumber\\
\end{eqnarray}

\begin{figure}[h]
\begin{center}
\subfloat{\includegraphics[width=0.16\textwidth]{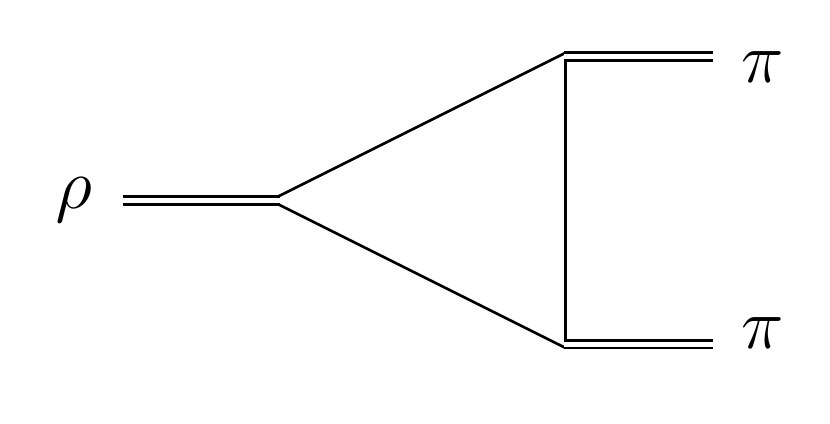}}
\subfloat{\includegraphics[width=0.16\textwidth]{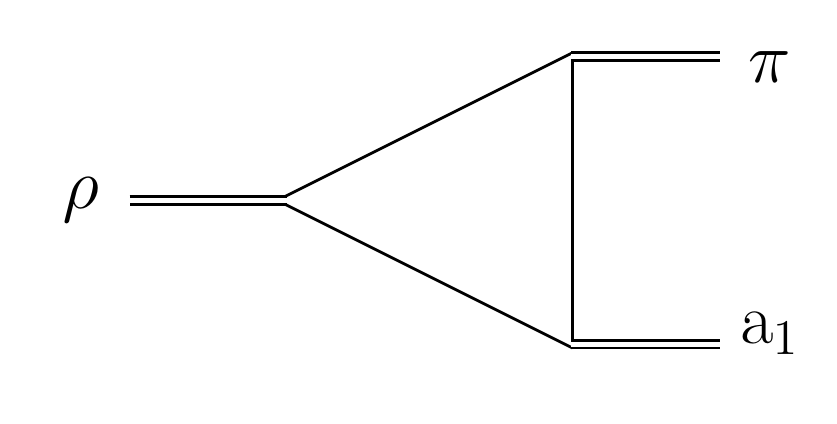}}
\subfloat{\includegraphics[width=0.16\textwidth]{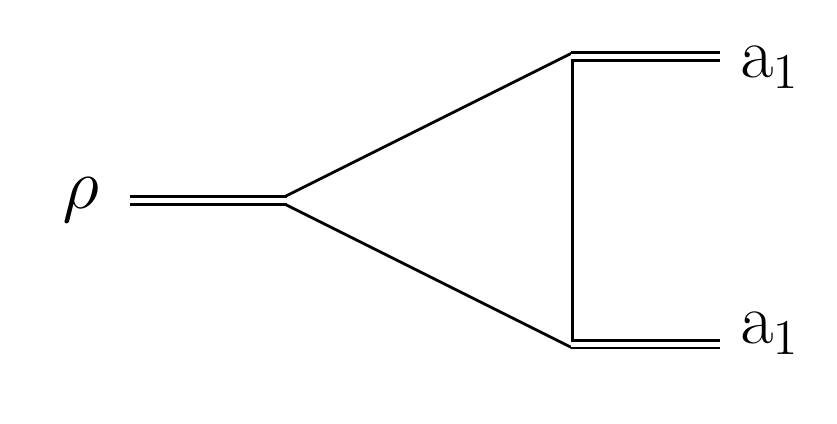}}\\
\subfloat{\includegraphics[width=0.16\textwidth]{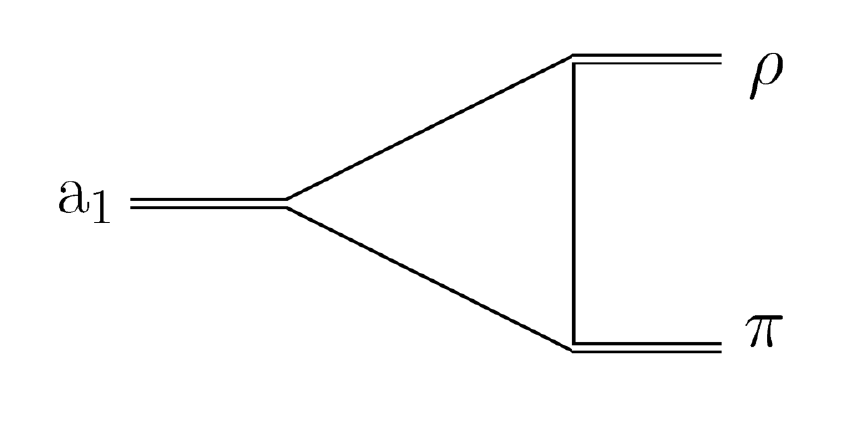}}
\subfloat{\includegraphics[width=0.16\textwidth]{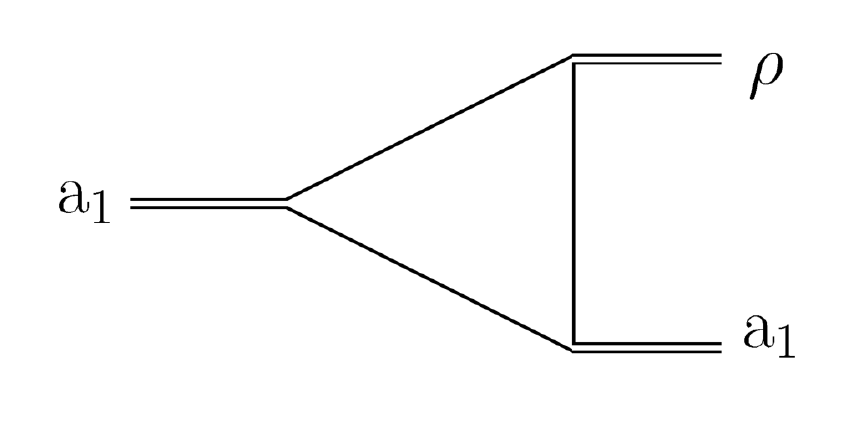}}
\end{center}
\caption{Diagrams contributing to $\rho$ and $a_1$ decays due to the $\pi-a_1$ mixing.}
\label{fig:diaggammas}
\end{figure}

The factors $g_{\rho\pi\pi}$, $g_{a\rho\pi}$ and $f_{a\rho\pi}$ are one-loop functions that arise from the expansion up to order three of the effective action. 
And, due to the $\pi-a_1$ mixing, receive contributions from the diagrams sketched in Fig.~\ref{fig:diaggammas}.
For the explicit forms of these functions and the temperature dependence of the widths, we refer the reader to Refs.~\cite{Villafane:2016ukb,Carlomagno:2019yvi}.


\section{Results}
\label{results}

To determine the relation between the perturbative QCD threshold $s_0$ and the trace of the Polyakov loop $\Phi$, we study the finite energy sum rules first at zero density, where chiral restoration and deconfinement occurs simultaneously.
Then we move into finite values of the chemical potential, where chiral symmetry is restored through a first order phase transition at a certain critical temperature whereas deconfinement has not yet been achieved, occurring the deconfinement transition at a higher temperature. 

Therefore, for large chemical potentials, the critical temperatures for the restoration of the chiral symmetry and deconfinement transition begin to separate. 
The region between them denotes a phase where the chiral symmetry is restored but quarks remains confined. 
This splitting is strongly dependent on the functional form of the Polyakov loop effective potential and also on the parameter $T_0$ entering in the PL potential~\cite{Carlomagno:2018tyk}. 
If we consider for this parameter an explicit dependence with $\mu$~\cite{Schaefer:2007pw,Ciminale:2007sr,Herbst:2010rf}, both transitions are always simultaneous, and therefore there is no such mixed phase.
\vspace*{0.5cm}

It should be mentioned that the FESR program was performed at one-loop order.
In the thermal perturbative QCD sector, only the leading one-loop contributions can be taken into account, since the problem of the appearance of two scales, i.e. the short-distance QCD scale and the critical temperature, remains unsolved.
However, at zero temperature it is possible to extend the calculations up to five-loop order.
The estimations at one-loop and five-loop order for $s_0$, $C_4 \langle \hat{\mathcal{O}}_{4}  \rangle$ and $C_6 \langle \hat{\mathcal{O}}_{6}  \rangle$ differ considerably~\cite{Ayala:2012ch,Dominguez:2012bs,Zhang:2012} (even with changes in the sign of the coefficients).
Therefore, only the thermal behavior of the condensates obtained through the higher order FESR should be taken into account.

The nlPNJL effective model includes six free parameters, namely the current quark mass $m$ and the coupling constants $G_S$, $G_V$, $G_0$, $G_5$ and $\kappa_p$~(see Eq.~(\ref{action})).
In addition, one has to determine the cutoffs $\Lambda_0$ and $\Lambda_1$ introduced in the form factors, Eq.~(\ref{regulators}).  

Through a fit to lQCD results quoted in Ref.~\cite{Parappilly:2005ei} for the functions $m(p)$ and $z(p)$, we obtain $\Lambda_0 = 1092 \pm 22$~MeV and $\Lambda_1 = 1173 \pm 60$~MeV.
Furthermore, by requiring that the model reproduces the value of $z(p = 0)$ and the  empirical values of three physical quantities, namely the masses of mesons $\pi$ and $\rho$ and the pion weak decay constant $f_{\pi}$, one can determine the model parameters quoted in Table~\ref{tab:param}.

\begin{table}[h]
\begin{center}
\begin{tabular*}{0.25\textwidth}{@{\extracolsep{\fill}} c c }
\hline
\hline
Parameter &  Value \\
\hline
$m$ [MeV] & 2.256  \\
$G_S$ [GeV$^2$] & 23.296 \\
$G_V$ [GeV$^2$] & 20.049  \\
$\kappa_p$ [GeV] & 4.265  \\
\hline
\hline
\end{tabular*}
\caption{\small{Model parameter values.}}
\label{tab:param}
\end{center}
\end{table}

Regarding the vector coupling constants $G_0$ and $G_5$, in the MFA only the former should be fixed~(see Eq.~(\ref{granp})).
Therefore, we will follow the prescription used in Ref~\cite{Contrera:2012wj}, parameterizing the isoscalar vector coupling as $G_0 = \eta\ G_V$.
Hence, the strength of the vector coupling can be evaluated by considering different values for $\eta$.

The influence of the vector coupling increases with the chemical potential.
At zero density, $\bar{\omega}$ vanishes for all temperatures and therefore the vector interactions do not contribute to the mean field thermodynamic potential.

Once the model parametrization is defined, one can calculate several meson properties at finite temperature and/or chemical potential.
For the numerical results of meson masses, decay constants, decay widths and other observables we refer the reader to Ref.~\cite{Carlomagno:2019yvi}.

The inputs used in the FESR calculated within the nlPNJL model are the masses, decay constants and decay widths of the $\pi$, $\rho$ and $a_1$ mesons.

\subsection*{FESR program at zero density}

At zero chemical potential, for the above set of parameters and for the polynomial Polyakov loop potential, Eq.~(\ref{upoly}), with $T_0 = 210$~MeV, we obtain through the corresponding susceptibilities, almost the same chiral and deconfinement critical temperatures $T_c = 202$~MeV (less than $3\%$ of difference), as expected, since at $\mu = 0$ chiral restoration and deconfinement takes place simultaneously as crossover phase transitions.
This behavior was verified by lQCD calculations~\cite{Bazavov:2016uvm}, in nlPNJL models~\cite{Contrera:2010kz,Carlomagno:2013ona,Carlomagno:2018tyk} and also obtained by finite energy sum rules~\cite{Ayala:2011vs}. 

Moreover, in Ref~\cite{Bazavov:2016uvm}, the deconfinement temperature defined at the peak of the entropy of a static quark (which is related to the Polyakov loop) is located at the same temperature, within errors, as the chiral susceptibility even at finite lattice spacing. 
In several works (see~\cite{Bazavov:2016uvm} and references therein), the deconfinement transition in lQCD with light dynamical quarks has been studied in terms of the inflection point of the renormalized Polyakov loop and fluctuations of conserved charges. 
Usually, these critical deconfinement temperatures are equal or larger than the restoring chiral transition critical temperature. 
In addition, these approaches have the disadvantage of being lattice scheme dependent and therefore the obtained values may differ considerably between them.

Furthermore, as it is discussed in Refs.~\cite{Braun:2007bx,Marhauser:2008fz,Herbst:2013ufa}, the strict comparison between our results and lattice data for the traced Polyakov loop has to be taken with some care, owing to the difference between the definitions of $\Phi$ in the continuum and on the lattice.
\vspace*{0.5cm}

From Eq.~(\ref{FESR}), for $N=1$ and $\mu=0$, we obtain the first FESR as function of the temperature $T$,
\begin{eqnarray}
&& 0 = 4 \pi \int_0^{s_0^{A,V}(T)} ds \, 
{\mbox{Im}}\ \Pi^{A,V}(s,T)\big\vert_{\mbox{\tiny{HAD}}} \nonumber \\
&& 
- \frac{4}{3}  \pi^2  T^2  - \int_0^{s_0^{A,V}(T)}ds \,\left[1 - 2\, n_F \left(\frac{\sqrt{s}}{2 T} \right) \right] \;,
\label{FESR1}
\end{eqnarray}
where $n_F(x)=1/(1+e^x)$ is the Fermi thermal function, and the spectral functions $\Pi^{A,V}|_{\mbox{\tiny{HAD}}}$ are given by Eqs.~(\ref{HADA}) and~(\ref{HADV}).
The continuum threshold $s_0^{A,V}$ can be calculated, as function of the temperature, by solving this equation with the corresponding spectral function.

\begin{figure}[h]
\begin{center}
\centering
\includegraphics[width=0.49\textwidth]{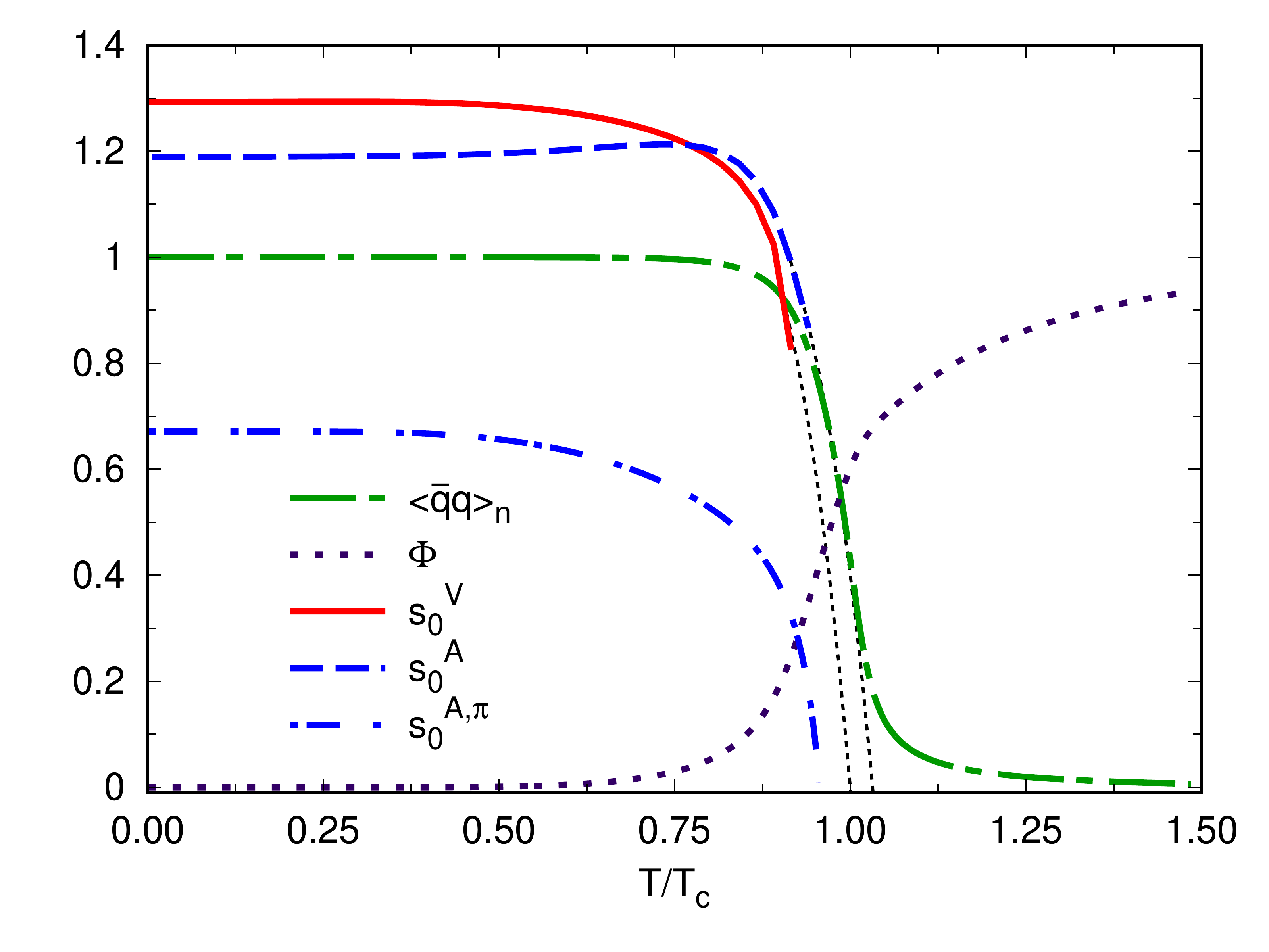}
\end{center}
\caption{Polyakov loop and normalized quark condensate in dotted and dash-dotted line, respectively, together with the continuum threshold $s_0$ for the vector (axial) channel in solid (dashed) line as function of $T/T_c$, with $T_c=202$~MeV.}
\label{fig:eses}
\end{figure}

In Fig.~\ref{fig:eses} we plot, as function of the reduced temperature $T/T_c$, the continuum threshold for the vector (axial) channel in solid (dashed) line, together with the trace of the PL and the quark condensate normalized by its value at $T=0$ in dotted and dash-dotted line, respectively. 

In addition, in dot-dashed line, we quote the continuum threshold for the axial-vector channel in the pion pole approximation $s_0^{A,\pi}$, its thermal behavior is equivalent to that found in our previous work, Ref.~\cite{Carlomagno:2016bpu}.

The FESR have solutions up to $T \sim 0.9\ T_c$, a temperature at which $s_0^{A,V}$ reaches its minimum.
A short extrapolation, denoted in the figure by a thin dotted line, should be understood for all results in the sequel.

As we expected for both channels, the PQCD threshold vanishes at critical temperatures $T_c^{V}=202$~MeV and $T_c^{A}=208$~MeV, located almost at the chiral critical temperature $T_c^{ch}=202$~MeV and the PL deconfinement temperature $T_c^{\Phi}=196$~MeV.

From the figure one can see that the thermal behavior of $s_0^A(T)$ and $s_0^V(T)$ close to $T_c$ is similar, even when the hadronic spectral functions are very different in these two cases.
This result is pointing to an approximate universality of the deconfinement transition in light-quark systems.

The higher order FESR, from where it is possible to estimate the thermal dependence of the gluon and four-quark condensate, can be analogously obtained from Eq.~(\ref{FESR}) with $N=2$ and $N=3$.
Both condensates show the expected behavior with a finite value at zero temperature, and decreasing monotonically as function of the temperature.

\begin{figure}[h]
\centering
\includegraphics[width=0.49\textwidth]{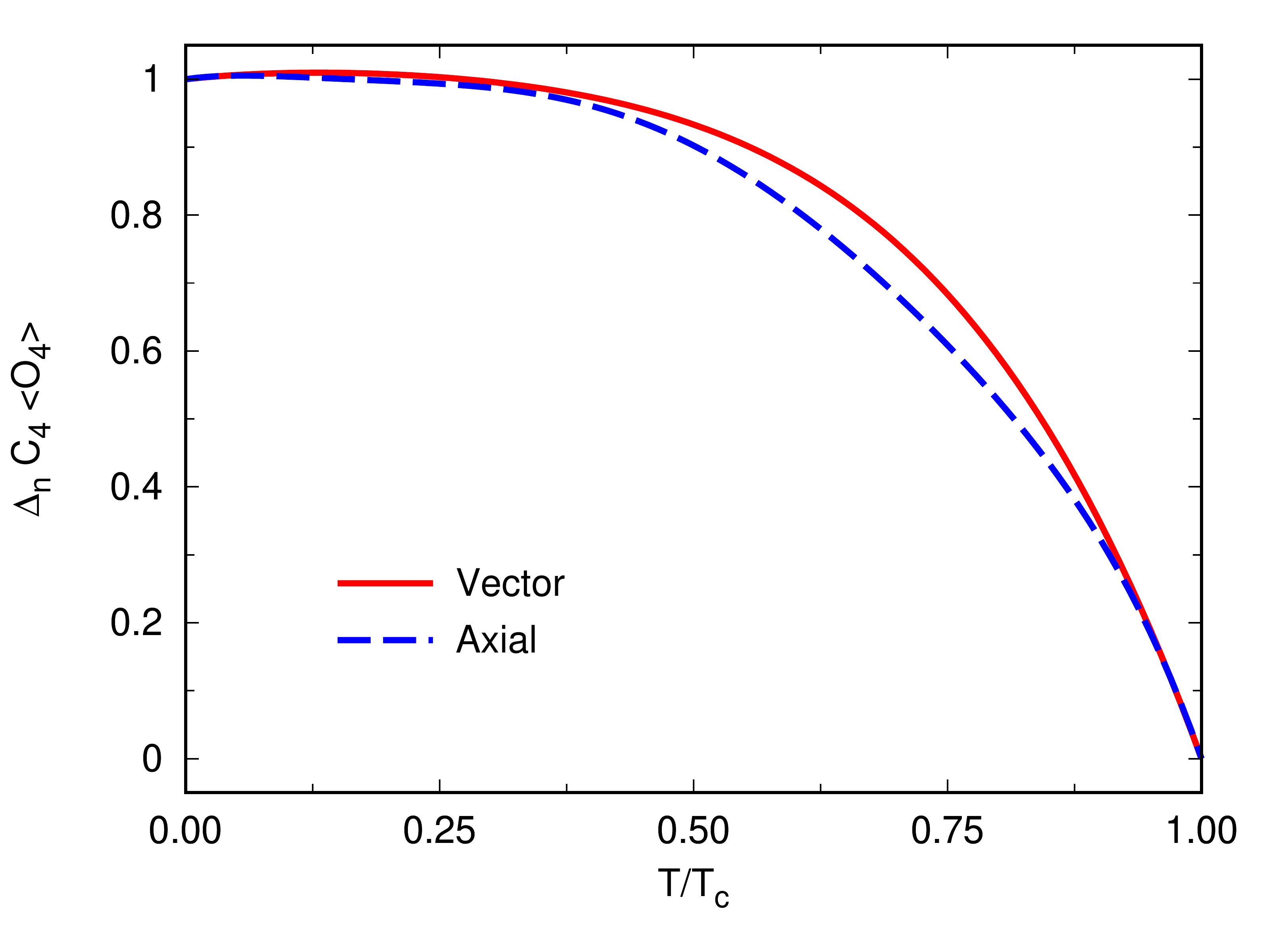}
\caption{Normalized $C_4 \langle \hat{\mathcal{O}}_{4} \rangle$, Eq.~(\ref{norm_c4o4}), for the axial and vector channel in solid and dashed line, respectively, as function of $T/T_c$. }
\label{fig:c4O4}
\end{figure}

In order to avoid the mentioned discrepancies at $T=0$ due to different loop-order calculations, we plot in Fig.~\ref{fig:c4O4} for the axial and vector channel in dashed and solid line, respectively, a normalized $C_4 \langle \hat{\mathcal{O}}_{4} \rangle$ as function of the reduced temperature, where we have defined such quantity as
\begin{equation}
\Delta_n C_4 \langle \hat{\mathcal{O}}_{4} \rangle =
\frac{C_4 \langle \hat{\mathcal{O}}_{4} \rangle(T/T_c) - C_4 \langle \hat{\mathcal{O}}_{4} \rangle(1)}
{C_4 \langle \hat{\mathcal{O}}_{4} \rangle(0) - C_4 \langle \hat{\mathcal{O}}_{4} \rangle(1)} \ .
\label{norm_c4o4}
\end{equation}

It can be seen from the figure that the thermal evolution of the gluon condensate is quite similar in both channels (as it is expected), even when the spectral functions, Eqs.~(\ref{HADA}) and~(\ref{HADV}), are completely different.

The thermal behavior of the meson masses directly affects the decay widths, since the kinematic condition in Eq.~(\ref{gammaofT}) tends to zero as $T$ increases and therefore, even when the phase space is increased due to the Bose enhancement, the width decreases~\cite{Carlomagno:2019yvi}.

For the process $\rho \rightarrow \pi \pi$, the decay width starts to drop above the chiral critical temperature, since beyond this temperature the $\pi$ mass grows faster than the $\rho$ mass.

For the $a_1$ decay, the width begins to diminish before the chiral critical temperature, vanishing close to $T_c$.
This is caused by the chiral partner mass degeneration. 
Near above $T_c$, vector mesons have approximately the same mass.

The physical decay process is $a_1 \rightarrow \pi \pi \pi$, however in our approach we are only considering the main channel of the partial decay $a_1 \rightarrow \rho \pi$.
Other partial widths contribute approximately with $40 \%$ of the total width~\cite{Tanabashi:2018oca}.
These processes, not considered here, will contribute to the total width and could modify the decreasing behavior of the $a_1$ width.

The consequence of this temperature dependence in $\Gamma_{a_1}$ is an small increment in the value of $s_0^A$ (see Fig.~\ref{fig:eses}).
This indicates that the non considered decay processes for the $a_1$ could be relevant for temperatures close to the critical temperature $T_c$.

Nevertheless, it should be noticed that although we are just considering the $a_1$ main decay channel, the general thermal dependence of the continuum threshold $s_0^A(T)$ not only is in agreement with other QCD sum rules results at one-loop order~\cite{Zhang:2012}, but also with the behavior found in this work for the vector channel.

\subsection*{FESR program at finite density}

In general, one can find regions in the QCD phase diagrams where chiral symmetry is either broken or restored through a crossover or a first order phase transition and regions where the system remains either in confined or deconfined states.

For relatively low densities, chiral restoration takes place as a smooth crossover, whereas for high values of chemical potential the order parameter has a discontinuity at a given critical temperature $T_c^\mu$ signaling a first order phase transition. 
This gap in the quark condensate induces also a jump in the trace of the PL, and the PL susceptibility present a divergent behavior at the chiral critical temperature.
Therefore, as in Ref~\cite{Contrera:2010kz}, when the chiral phase transition is first order, we define the deconfinement critical temperature $T_{\Phi}^\mu$ requiring that $\Phi=0.4$, which could be taken as large enough to denote deconfinement.

The value of $\Phi$ at both sides of the discontinuity indicates, for a chiral symmetric state, whether the system remains confined or not. 
The region where the chiral symmetry is restored but the quarks and gluons remain confined, is usually referred as the quarkyonic phase~\cite{McLerran:2007qj,McLerran:2008ua,Abuki:2008nm}.

If we move, in the $T-\mu$ plane, along the first order phase transition curve, the critical temperature rises from zero up to a critical endpoint (CEP) temperature $T_{CEP}$, while the critical chemical potential decreases from its value at zero temperature $\mu_c$ to a critical endpoint chemical potential $\mu_{CEP}$. 
Beyond this point, the chiral restoration phase transition proceeds as a crossover. 

To determine these temperatures and densities we need to fix the value of the coefficient $\eta$ in the definition of $G_0$.
Here, as in Ref.~\cite{Carlomagno:2019yvi}, we choose $\eta = 0$ and $\eta = 0.3$ as representative cases.
Since the former leads to a mean field theory without vector interactions, and the later provides the best phenomenological agreement with other effective models (see Ref.~\cite{Contrera:2012wj} and references therein).

\begin{figure}[H]
\begin{center}
\centering
\includegraphics[width=0.49\textwidth]{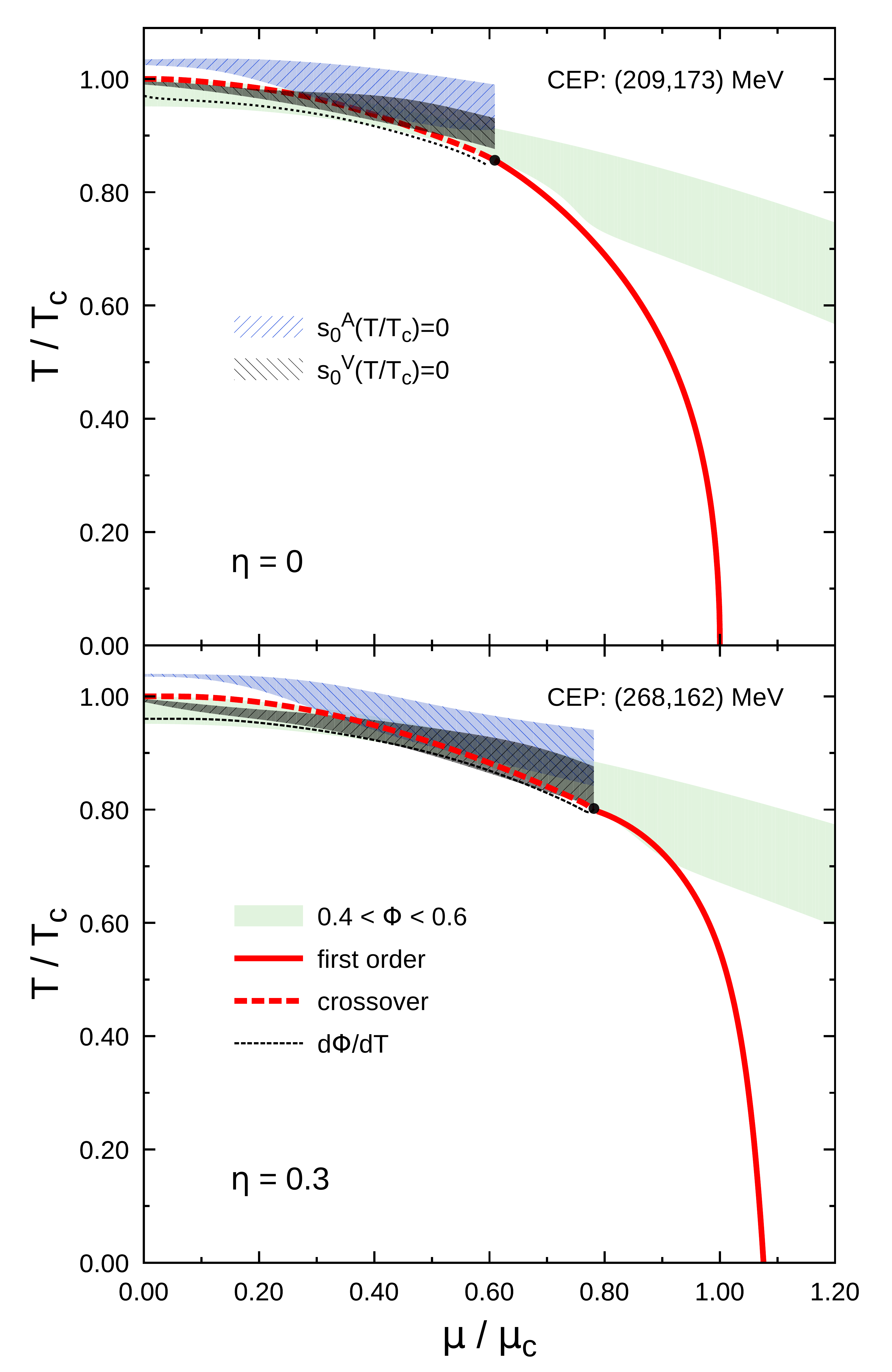}
\end{center}
\caption{QCD phase diagrams for $\eta=0$ and $\eta=0.3$ in the upper and lower panel, respectively~\cite{Carlomagno:2019yvi}, together with the $(\mu,T)$ coordinates and position of the CEP.
The $s_0^{A,V}$ associated deconfinement temperatures are represented by the pattern filled areas.}
\label{fig:qcdpd}
\end{figure}

In Fig.~\ref{fig:qcdpd} we plot the phase diagram for a mean field theory with and without vector interactions.
Specifically, in the upper and lower panel we quote the reduced critical temperatures as function of the reduced chemical potential for $\eta=0$ and $\eta=0.3$, respectively.

In the crossover region the deconfinement temperatures are determined with the PL susceptibility, whereas for $\mu>\mu_{CEP}$ are obtained by requiring that $0.4<\Phi<0.6$, quoted in dotted line and with the color shaded area, respectively.
In addition, the chiral critical temperatures are represented by the solid and dashed line, for the first order and crossover phase transition curves, respectively.
Finally, the dot indicates the position of the critical endpoint.
These phase diagrams were taken from Ref.~\cite{Carlomagno:2019yvi}.

At zero density the FESR have solutions up to $0.9\ T_c$ where the continuum threshold reaches its minimum.
Therefore, to define the value of $T$ where $s_0^{A,V}=0$, it is usual to perform a short extrapolation.

However, in the crossover region, when $\mu$ increases the FESR equations stop having solutions at lower temperatures and consequently the extrapolation is no longer well defined.
This led us to set a range of critical temperatures $T^\mu_{s_0^{A,V}}$, for each value of $\mu$, that enlarges as the density grows.

We quote this, in both panels of Fig.~\ref{fig:qcdpd}, with pattern filled areas.
The critical temperatures within those bands, correspond to a family of extrapolations of the obtained results, for both channels, with similar $\chi^2$.

Despite this broadening, it can be seen that the density dependence of $T^\mu_{s_0^{A,V}}$ is equivalent to that found for $T^\mu_{\Phi}$ and $T^\mu_{c}$, showing that the continuum threshold and the traced Polyakov loop provide analogous information about the QCD deconfinement transition in a wide range scenario.

\begin{figure}[h]
\begin{center}
\centering
\includegraphics[width=0.49\textwidth]{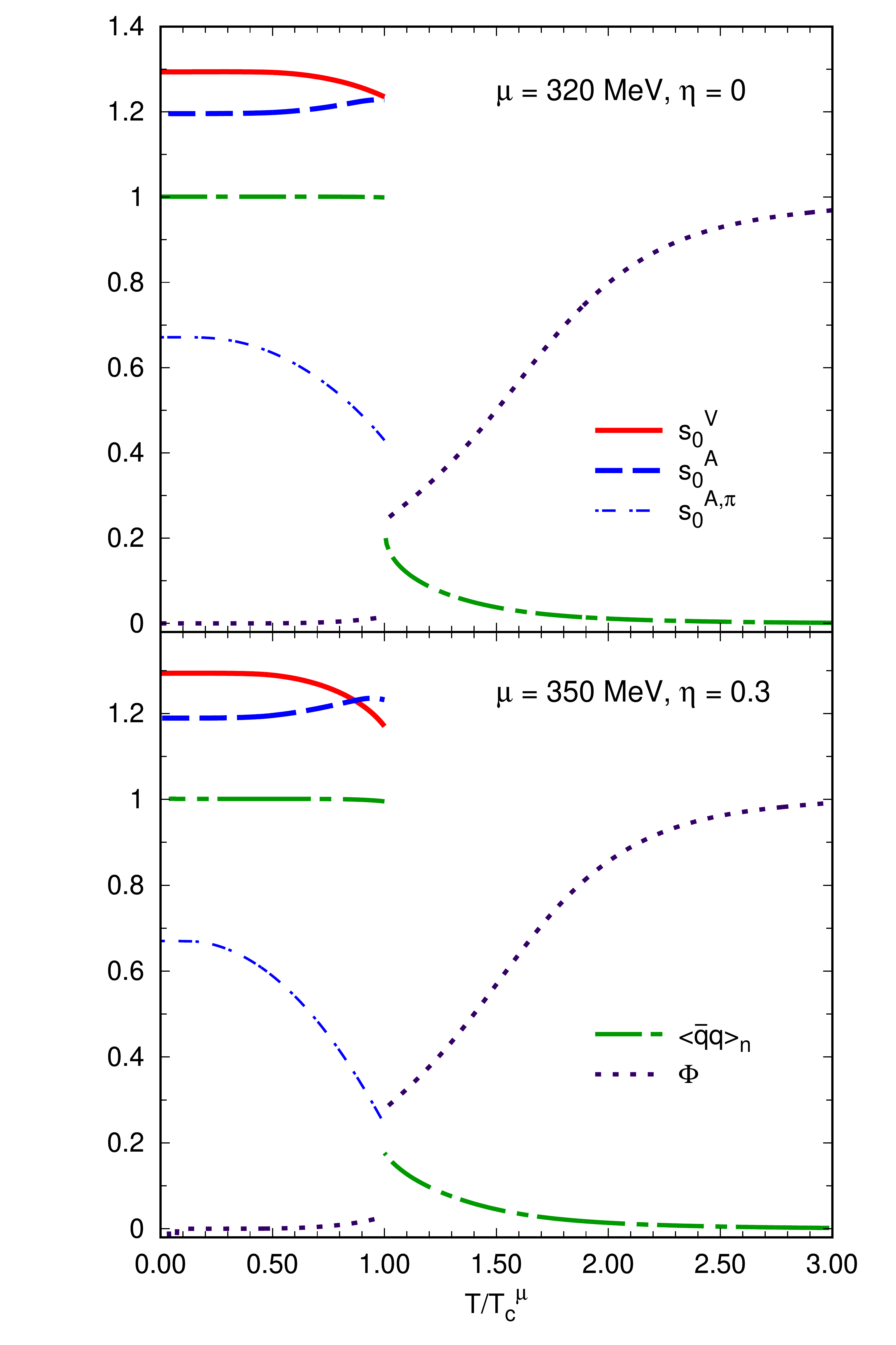}
\end{center}
\caption{Polyakov loop and normalized quark condensate in dotted and dash-dotted line, together with the continuum threshold $s_0$ for the vector (axial) channel in solid (dashed) line, for $\eta = 0$ and $\eta = 0.3$, in the upper and lower panel, respectively.}
\label{fig:eses_mu}
\end{figure}

On the other hand, for chemical potentials larger than the $\mu_{CEP}$, the thermal dependence of the continuum threshold in the available temperature range, does not allow to extrapolate the $s_0$ curve beyond the last obtained value, usually located at $T  \gtrsim T_c^\mu$ (see Fig.~\ref{fig:eses_mu}).

In a finite density scenario, when $\mu > \mu_{CEP}$, the two transitions takes places separately at different critical temperatures, and therefore provides unique conditions to identify the phenomenological equivalence between $s_0$ and $\Phi$ as deconfinement order parameters.

Particularly, we choose $\mu=320$~MeV and $\mu=350$~MeV for $\eta = 0$ and $\eta = 0.3$, respectively.
Since for these values of chemical potential, chiral and deconfinement critical temperatures are separated by approximately $25-30$~MeV.

In Table~\ref{tab:cep} we summarize, for these two scenarios, the critical temperatures, critical chemical potentials and the CEP coordinates.
\begin{table}[H]
\begin{center}
\begin{tabular*}{0.3\textwidth}{@{\extracolsep{\fill}} ccc }
\hline 
\hline 
$G_0$ & 0 & 0.3\ $G_V$ \\
\hline 
$T_{\rm CEP}$ [MeV] & 173 & 162 \\
$\mu_{\rm CEP}$ [MeV] & 209 & 268 \\
\hline 
$T_c$ [MeV]  & 202 & 202 \\
$\mu_c$ [MeV]  & 343 & 369 \\
\hline
  & $(\mu=320)$ & $(\mu=350)$ \\
$T_c^\mu$ [MeV] & 102 & 108 \\
$T_{\Phi}^\mu$ [MeV] & 136 & 134 \\
\hline 
\hline 
\end{tabular*}
\caption{\small{CEP coordinates and critical temperatures and densities for both cases of vector strength.}}
\label{tab:cep}
\end{center}
\end{table}

In the upper and lower panel of Fig.~\ref{fig:eses_mu} we plot for $\eta = 0$ and $\eta = 0.3$, respectively, the continuum threshold for the vector (axial) channel in solid (dashed) line, together with the trace of the PL and the normalized quark condensate in dotted and dash-dotted line, respectively. 

As before, we also quote in dot-dashed line, the continuum threshold for the axial-vector channel in the pion pole approximation, $s_0^{A,\pi}$.

In both situations, $\eta=0$ and $\eta=0.3$, we see that for bigger densities than the critical end point chemical potential, the thermal equation has not solution beyond the critical temperature $T_c^\mu$. 
The continuum threshold, for both channels, stops with a finite value at this temperature, signaling that the system continues in a confined state.

As in the zero density case $s_0^A$ has a slight increase before decreasing, consequence of the approximation used for the $a_1$ width decay.

Regarding the Polyakov loop, the value of $\Phi$ at both sides of $T/T_c^\mu=1$ indicates that the system, at this temperature, remains in a confined state, even when the chiral symmetry has been restored. 

In this way, we see that the Polyakov loop and the continuum threshold provide the same information. 
When the chiral symmetry is restored, $s_0$ and $\Phi$ show that we are still in a confined phase. This characterizes the occurrence of a quarkyonic phase.

\section{Summary and conclusions}
\label{finale}

Along this article, as in our previous work~\cite{Carlomagno:2016bpu}, we compare the behavior of two vastly used phenomenological order parameters for the deconfinement transition: the continuum threshold $s_0$ and the trace of the Polyakov loop $\Phi$.

In Ref.~\cite{Carlomagno:2016bpu} we study the finite energy sum rules for the axial-vector current correlator saturating the spectral function with the pion pole approximation.
Here, we have extended that analysis in two complementary directions: we improve the approximation for the axial spectral function including the $a_1$ resonance, and we consider the vector current correlator assuming $\rho$-meson saturation for the spectral function. 

In this way, both on the side of the FESR formalism and the nlPNJL model we have considered the best possible phenomenological approach.
Since there are no further possible corrections to the spectral function or to the effective model in the light quark sector, our results seem to be strong and conclusive.
 
The input parameters used in the FESR, namely the masses, decay constants and decay widths for the $\pi$, $\rho$ and $a_1$ mesons, were obtained from a SU(2) PNJL model with nonlocal vector and axial-vector interactions~\cite{Carlomagno:2019yvi}.

At zero density, we determine that the continuum threshold vanishes, for both channels, at approximate the same temperature where the Polyakov susceptibility has its maximum value. 

At finite density, beyond the critical end point chemical potential, we found that for both deconfinement parameters, the system remains in its confined phase even when the chiral symmetry is restored.  

Therefore, based in our previous study and the results obtained here, we can conclude that both quantities, $s_0(T,\mu)$ and $\Phi(T,\mu)$, provide the same kind of physical information about the QCD deconfinement transition.


\section*{Acknowledgements}
Support for this work has been received in part by National University of La Plata (Argentina), Project No.\ X824 and Subsidio para Estad\'ias 2018. 
M. L. acknowledge support from FONDECYT (Chile) under grants No. 1170107 and No. 1190192 and in addition, M. L. acknowledges support from CONICYT PIA/BASAL (Chile) grant No. FB0821.




\end{document}